\documentstyle[12pt]{article}
\begin{document}
\font\bss=cmr12 scaled\magstep 0
\title{Tachyonic Instability and Darboux Transformation}
\author{ A.V. Yurov
\small\\ Theoretical Physics Department,
\small\\ Kaliningrad State University, Russia,  yurov@freemail.ru}
\date {}
\renewcommand{\abstractname}{\small Abstract}
\maketitle
\maketitle
\begin{abstract}
Using Darboux transformation one can construct infinite family of potentials which lead 
to the flat spectrum of scalar field fluctuations with arbitrary 
multiple precision, and, at the same time, with "essentially blue" spectrum of perturbations 
of metric. Besides, we describe reconstruction problem: find classical potential 
$V(\phi)$ starting from the known "one-loop potential" $u(t)=d^2V(\phi(t))/d\phi(t)^2$.
\end{abstract}
\thispagestyle{empty}
\medskip

\section{Introduction.}
In [1] a new mechanism for the generation density perturbation by tachyonic instability 
is discussed. If we assume that
\newline
\newline
1. the universe was flat and homogeneous;
\newline
2. Hubble constant is small during the process generation of the perturbation
\newline
\newline
then the equation for the quantum fluctuations $\delta\phi_k(t) e^{-i\bf{kr}}$ of scalar field 
is
$$
\delta\ddot\phi_k+\left(k^2+u(t)\right)\delta\phi_k=0,
\eqno(1)
$$
where $u(t)=d^2V(\phi(t))/d\phi(t)^2$ (we call $u(t)$ as "one-loop potential"), 
$V(\phi)$ is classical potential, $\phi(t)$ is a 
homogeneous solution of classical equation of motion. To calculate spectrum of density 
one need to find solution of (1) with the asymptotic behavior 
$\delta\phi_k(t\to+\infty)=e^{ikt}/\sqrt{2k}$. Then density fluctuations at 
moment $t_0$ are given by 
$$
\delta_k\sim H\left(\frac{k}{2\pi}\right)^{3/2}\frac{\delta\phi_k(t_0)}{\delta\phi_0(t_0)},
\qquad
n_s=1+\frac{d\log\mid\delta_k\mid^2}{d\log k},
\eqno(2)
$$
where $\delta\phi_0$ is solution of the (1) with $k=0$, $n_s$ is the spectral 
index deciding the "color" of spectrum. It is easy to see that $\delta\phi_0$ is both 
$k=0$ solution of the (1) and the time derivative of the field $\phi(t)$: 
$\delta\phi_0(t)={\dot{\phi}(t)}$ so it's dimensions is $[\delta\phi_0]=$GeV${}^2$ whereas 
for the $\delta\phi_k(t)$ we choose the normalization factor to have 
$[\delta\phi_k]=$GeV${}^{1/2}$ (in the units with $c=\hbar=1$).

There are just two  usually  studied  potentials: power-law potential 
$V(\phi)=-\lambda\phi^n/M^{n-4}$, ($M$ is the some constant, 
$\lambda>0$~\footnote{The potential $V(\phi)<0$ so one needs to add to $V(\phi)$ some 
other terms to stabilize the motion. But it doesn't matter in this paper which is 
devoted to tachyonic instability.})
and "ekpyrotic" or "cyclic" potential $V(\phi)=-V_0 e^{-\phi/M}$ [2, 3]. Both of them 
leads to the equation (1) with $u(t)=-\mu^2/t^2$. For the power-law potentials 
$\mu^2=\mu^2(n)=2n(n-1)/(n-2)^2$ and it easy to see 
that all quantities $\mu^2(n)>0$ can be obtained by appropriate choosing of $n$. For example, 
if $n>2$ then $\mu^2(n)\in (2,+\infty)$; if $n\in [1,4/3]$ then $\mu^2(n)\in [0,2]$ and for the 
$n\in (-\infty,0]$ we have $\mu^2(n)\in [0,2)$. For the "cyclic" potential $\mu^2=2$.

In these both cases the equation (1) can be solved via Hankel function. The aim of this work 
to call attention to simple but effective method to construct infinite set of one-loop 
potentials with two properties:
\newline
\newline
1. All this one-loop potentials are integrable. This allow one to find $\delta\phi_k$ and $\delta\phi_0$ exactly to calculate $\delta_k$. 
\newline
2. These one-loop potentials   lead to almost flat spectrum. 
\newline
\newline

Although, our approach get us a family "one-loop" potentials $u(t)$ one can 
reconstruct initial classical potential $V(\phi)$ starting out from the $u(t)$. 
We do it in the Sec. 3. In this next  we also discuss the compatibility one of 
new potential with inflation. One show that this potential can lead 
to inflation but for very specific initial condition. This fact is at one with  
analysis in [1].  

Our results about the flat spectrum are valid only for the spectrum of the scalar field and, 
to all appearances, 
don't valid for the spectrum of density perturbations. In Sec. 4, we show that using Darboux transformations, 
one can construct infinite set exact soluble cosmologies with flat spectrum of the scalar field 
perturbations and, at the same time, with  "essentially nonflat", blue spectrum of 
perturbations of metric. 

 \section{Flat spectrum via Darboux transformation.}
Let consider the equations 
$$
\delta\ddot{\phi}_k+(k^2+u(t))\delta\phi_k=
\delta\ddot{\psi}_{\kappa}+(-\kappa^2+u(t))\delta\psi_{\kappa}=0,
\eqno(3)
$$
where $u(t)=-2/t^2$, $k$ and $\kappa$ are real numbers. One choose the solution of the first 
equation (3) as ($t>0$). 
$$
\delta\phi_k=N\left(1+\frac{i}{kt}\right)e^{ikt}.
$$
This is the particular case of the Hankel functions [1]. 
To calculate spectrum one need to choose the normalization factor $N=(2k)^{-1/2}$. It is shown 
in [1] that in this case the spectrum of density fluctuations will be exactly flat, $n_s=1$.
It's clear that $\delta\phi_0$  has the form 
$$
\delta\phi_0=c_+t^2+\frac{c_-}{t},
\eqno(4)
$$
$c_{\pm}$ are constants of integration. 

The general solution of the second equation from (3) is 
$$
\delta\psi_{\kappa}=C_+\left(1-\frac{1}{\kappa t}\right)e^{\kappa t}+
C_-\left(1+\frac{1}{\kappa t}\right)e^{-\kappa t},
\eqno(5)
$$
with some constants $C_{\pm}$. We introduce the Darboux transformation for the (3) as
$$
\begin{array}{l}
\displaystyle{
\delta\phi_k(t)\to\delta\phi^{(1)}_k(t;\kappa)\equiv\delta\phi^{(1)}_k=
\frac{\delta\dot{\phi_k}\delta\psi_{\kappa}-\delta\phi_k\delta\dot{\psi_{\kappa}}}
{\delta\psi_{\kappa}},}\\
\\
\displaystyle{
u(t)\to u^{(1)}(t;\kappa)\equiv u^{(1)}=u+2\frac{d^2\log\delta\psi_{\kappa}}{dt^2}.}
\end{array}
\eqno(6)
$$
It's easy to see that new function $\delta\phi^{(1)}_k$ is solution of "dressed" equation 
$$
\delta{\ddot{\phi}}^{(1)}_k+\left(k^2+u^{(1)}\right)\delta\phi^{(1)}_k=0,
\eqno(7)
$$
if $\delta\phi_k$ and $\delta\psi_{\kappa}$ are solutions of the (3). We call 
$\delta\psi_{\kappa}$ as {\bf prop function}. We can choose the prop function as any 
solution of the equation (3), for example as $\delta\phi_{\tilde k}$ with 
$\tilde k\ne k$,  or as $\delta\phi_0$, 
but it is useful to choose one as (5). In this case new 
(integrable~\footnote{This is because 
one can obtain all solutions of the (7) starting out from the known solutions of the 
eq. (3). This 
why Darboux transformation is power tool to construct many, if not all, exact 
soluble potential 
in one dimensional quantum mechanics [5].}) potential $u^{(1)}$ has the form,
$$
u^{(1)}=-2\kappa^2\frac{A^2e^{2\kappa t}+B^2 e^{-2\kappa t}-4AB(\kappa t)^2-2AB}
{\left(A(\kappa t-1)e^{\kappa t}+B(\kappa t+1)e^{-\kappa t}\right)^2},
\eqno(8)
$$
so when $t\to+\infty$ then
$$
u^{(1)}\to -\frac{2}{t^2}-\frac{4}{\kappa t^3}\to u(t).
$$
It mean that for the enough large $t_0$,  $u^{(1)}(t_0)\sim u(t_0)$ and therefore, the spectrum 
of fluctuations at this moment is the same both $u(t_0)$ and $u^{(1)}(t_0)$. 

To show this we choose the normalization factor of initial $\delta\phi_k$ as 
$N=1/(ik-\kappa)\sqrt{2k}$,
to obtain the good asymptotic behavior for the $\delta\phi^{(1)}_k$: 
$\delta\phi^{(1)}_k(t\to+\infty;\kappa)=e^{ikt}/\sqrt{2k}$. We suppose that
$$
0<kt_0\ll 1\ll \kappa t_0,
\eqno(9)
$$
so we are interesting of long wavelength fluctuations with wavelength 
$\lambda\sim 1/k\gg 1/\kappa$, at the moment $1/\kappa\ll t_0\ll 1/k$.

Using (6), (9) we get 
$$
\left|\delta\phi^{(1)}_k(t_0;\kappa)\right|^2\sim\frac{k}{2\left(k^2+\kappa^2\right)}
\left(2+\frac{1}{(kt_0)^4}+\frac{2}{(kt_0)^3}\right).
$$

To obtain $\delta^{(1)}_k$ one need to substitute $\delta\phi^{(1)}_k(t_0;\kappa)$ and 
$\delta\phi^{(1)}_0(t_0;\kappa)$ into the (2). The function $\delta\phi^{(1)}_0(t_0;\kappa)$ 
can be obtained from the (6) by the substitution $\delta\phi_k\to\delta\phi_0$ from the (4). 
The only $k$ dependence is in the $\delta\phi^{(1)}_k(t_0;\kappa)$, so we omit the 
calculation of $\delta\phi^{(1)}_0(t_0;\kappa)$. It's clear that in main order 
($\kappa\gg k$, $kt_0\ll 1$) the amplitude $\mid\delta^{(1)}_k\mid$ does not depend on $k$, 
so we  get a flat spectrum. As in the [1], we obtain this result without any brane-string 
physics. 

To consider small deviation from flat spectrum we write 
$$
\mid\delta\phi^{(1)}_k\mid\sim \frac{1}{\kappa \sqrt{2}t_0^2k^{3/2}}\left(1+kt_0\right).
$$
Using (2) we get 
$$
n_s=2\left(2+\frac{k}{\mid\delta\phi^{(1)}_k(t_0)\mid}\frac{d\mid\delta\phi^{(1)}_k(t_0)\mid}
{dk}\right)=1-kt_0,
$$
so if $kt_0\to 0$ then $n_s\to 1$ and $n_s<1$. We get red spectrum but for the small $kt_0$ 
(large wavelengths) the deviations from the flat spectrum are small. As suggested 
by observation [4] we need $\mid n_s-1\mid <0.1$, so $\mid kt_0\mid<0.1$.

A single act of dressing (6) can be iterated $n$ times [5]. As a result one get 
$$
u^{(n)}=u+2\frac{d^2}{dt^2}\log\det\left(\begin{array}{cccc}
d^{n-1}\delta\psi_{\kappa_n}/dt^{n-1}&d^{n-2}\delta\psi_{\kappa_n}/dt^{n-2}&...&\delta\psi_{\kappa_n}\\
d^{n-1}\delta\psi_{\kappa_{n-1}}/dt^{n-1}&d^{n-2}\delta\psi_{\kappa_{n-1}}/dt^{n-2}&...&\delta\psi_{\kappa_{n-1}}\\
.\\
.\\
d^{n-1}\delta\psi_{\kappa_1}/dt^{n-1}&d^{n-2}\delta\psi_{\kappa_1}/dt^{n-2}&...&\delta\psi_{\kappa_1}\\
\end{array}
\right),
$$
where $\delta\psi_{\kappa_m}$ is solution of the equation (3) 
with $\kappa=\kappa_m$, $m=1,..,n$. Choosing $\delta\psi_{\kappa_m}$ at the form (5) we 
get at the moment $t_0\gg1/K$, with $K=min\left\{\kappa_m\right\}$,
$$
u^{(n)}(t_0)\sim u(t_0)=-2/t_0^2,
$$
therefore the one-loop potential $u^{(n)}$ lead to the flat spectrum just as $u^{(1)}$ (8). 

It is interesting to obtaine the classical potential $V(\phi)$ which lead to "one-loop potential" 
(8). We'll do it in the next section. 
\section {Reconstruction of potential $V(\phi)$.}

To reconstruct $V(\phi)$~\footnote{The general fomulation 
of the reconstruction problem take place in our work [6]}we start with equation of motion for the field $\phi=\phi(t)$,
$$
\ddot\phi=-V'(\phi).
\eqno(10)
$$
Introducing new variable $\eta(t)=\dot\phi$ we get 
$$
{\ddot\eta}+V''(t)\eta=0,
\eqno(11)
$$
 Comparing with (1) one conclude that 
$\eta(t)= \delta\phi_0(t)$. Solving (11) we find $V(\phi)$ 
in the parametric form 
$$
V(t)=\rho-\frac{1}{2}\eta^2(t),\qquad 
\phi(t)=\phi_0+\int dt\,\eta(t),
\eqno(12)
$$
where $\rho$=const is energy density and $\phi_0$=const is initial value of $\phi(t)$. 

Let illustrate this simple method for the 
$V''=-\mu^2/t^2$. If $\mu^2\ne 2$ then the general solution of the (11) is 
$$
\eta(t)=\sqrt{t}\left(C_+t^{\beta}+C_-t^{-\beta}\right),
\eqno(13)
$$
where 
$$
\beta=\frac{1}{2}\sqrt{1+4\mu^2}=\frac{1}{2}\left|\frac{3n-2}{n-2}\right|.
$$
The quantity $\beta=3/2$ (flat spectrum) when $n\to \infty$ (it lead ones to 
the cyclic potential) 
and for the power-low potential with $n=4/3$. Thus, for the $\mu^2(n)\ne 2$ one have
$$
V(t)=\rho-\frac{1}{2}\left(C_+t^2+\frac{C_-}{t}\right)^2,
\qquad
\phi(t)=\phi_0+2t^{2/3}\left(\frac{C_+}{3+2\beta}t^{\beta}+\frac{C_-}{3-2\beta}t^{-\beta}\right),
\eqno(14)
$$
and for the $\mu^2=2$
$$
V(t)=\rho-\frac{1}{2}\left(C_+t^2+\frac{C_-}{t}\right)^2,
\qquad
\phi(t)=\phi_0+\frac{C_+}{3}t^3+C_-\log t.
\eqno(15)
$$

In a case of general position one can't find $V(\phi)$ in non-parametric form but we 
can do it if one of constants $C_{\pm}$ is zero.  Let $C_-=0$, then 
$$
V_+(\phi)=\rho-g^2\left(\phi-\phi_0\right)^{2(2\beta+1)/(2\beta+3)},\qquad 
g^2=\left(\frac{C_+^4\left(2\beta+3\right)^{2(2\beta+3)}}{2^{6\beta+5}}\right)^{1/(2\beta+3)}.
$$
We choose  $\rho=0$, $\phi_0=0$. Then 
$$
\begin{array}{l}
V_+(\phi)=-g^2\phi^{N_1(n)}, \qquad n>2,\\
\\
V_+(\phi)=-g^2\phi^n,\qquad n\in[1,4/3],\\
\\
V_+(\phi)=-g^2\phi^{N_2(\mid n\mid)},\qquad n\in (-\infty,0],
\end{array}
\eqno(16)
$$
where 
$$
N_1(n)=\frac{4(n-1)}{3n-4},\qquad 
N_2(\mid n\mid)=\frac{4(\mid n\mid +1)}{3\mid n\mid+4}.
$$
It is valid $4/3<N_1(n)<2$ and $1\le N_2(n)<4/3$.  In this case (which is 
valid both for the (14) and (15)) the absolute value of $\phi(t)$ grows as $t\to \infty$. 
To obtain decreasing ones we choose $C_+=0$. At last 
we get for the (14)
$$
\begin{array}{l}
V_-(\phi)=-g^2\phi^n,\qquad n>0\\
\\
V_-(\phi)=-g^2\phi^{N_1(n)}, \qquad n\in[1,4/3],\\
\\
V_-(\phi)=-g^2\phi^{N_2(\mid n\mid)},\qquad n\in (-\infty,0],
\end{array}
\eqno(17)
$$
and for the (15)
$$
V_-(\phi)=\rho-\frac{C_-^2}{2}e^{-2(\phi-\phi_0)/2}.
\eqno(18)
$$
The last example is nothing but "cyclic" potential.

All known potentials (16), (17), (18) are particular cases of (14), (15) which can 
be reconstructed in non-parametric representation. 
In a case of general position, these potentials can be obtained in parametric form only. 
Let consider (14). If $C_+<0$ and $C_-<0$ then function $\phi(t)$ is monotone  
decreasing function and $V(\phi)$ is single-valued function on $\phi$ with maximum in  
$$
\phi_*=\phi_0-\frac{\mid C_-\mid}{3}
\left(\frac{1}{2}+\log\frac{\mid C_-\mid}{2\mid C_+\mid}\right),\qquad
V(\phi_*)=\rho-\frac{9}{2}\left(\frac{\mid C_+\mid C_-^2}{4}\right)^{2/3}.
$$
This decreasing potential has asymptotic behavior
$$
V(\phi)\to -\frac{C_-^2}{2}e^{2\phi/\mid C_-\mid},\qquad as\qquad \phi\to+\infty \qquad 
(t\to 0),
$$ 
and 
$$
V(\phi)\to -\frac{C_+^{2/3}}{2}\left(3\phi\right)^{4/3},\qquad as\qquad \phi\to -\infty \qquad
(t\to+\infty).
$$
If, $C_+C_-<0$ then the function $\phi(t)$ has one extreme point and $V(\phi)$ is two-digit 
function on $\phi$. 

Using this method one can reconstruct the classical potential $V(\phi)$ starting out with the 
"one-loop potential" $u^{(1)}$ (8).    Dressing 
$\delta\phi_0$ (4) and choosing $C_+=C_-=1$, $c_-=0$ we get 
$$
\begin{array}{l}
\displaystyle{
u^{(1)}(y)=\frac{d^2V^{(1)}(\phi)}{d\phi^2}=-\frac{2\kappa\left(\sinh^2y-y^2\right)}{\left(y\cosh y-
\sinh y\right)^2} ,}\\
\\
\displaystyle{
\eta(y)=\frac{y\left(-3y\cosh y+(3+y^2)\sinh y\right)}{y\cosh y-\sinh y},\qquad y=\kappa t,}
\end{array}
\eqno(19)
$$
so using (12) one get 
 the potential $V(\phi)$ in parametric form. It easy to see that 
$$
V(\phi)\sim -\frac{1}{2}\left(3\kappa\phi\right)^{4/3},\qquad as \qquad t\to+\infty,
$$
and 
$$
V(\phi)\sim\rho-\frac{4}{\sqrt{5}}\left(\kappa(\phi-\phi_0)\right)^{3/2},\qquad
as\qquad t\to 0,
$$
therefore $\rho$ is initial value of $V(\phi)$. Then $dV/d\phi=2\kappa{\dot\eta}\ne 0$, 
so $V(\phi)$ is monotone  function on $\phi$. And last but not least, $\delta\phi_0$ 
is monotone function on time therefore $V(\phi)$ is single-valued function on $\phi$. 

Choosing another  $c_{\pm}$ one can find a family of classical potentials with the 
same quantum potentials $V''$ and which lead to the same spectrum. 

It is interesting to consider the last potential (which can be obtained from the (19)) from the 
inflation standpoint. To obtain inflation one 
suppose that ${\dot\phi}^2/2\ll \mid V(\phi)\mid$ so $\rho\gg \eta^2/2$ and during 
inflation $V(\phi)\sim\rho$ (we suppose  $\rho>0$). Therefore $H^2=8\pi\rho/3M_p^2=$const. 
But $k^2>H^2$ [1]. This is because we neglect the term $3H\delta{\dot \phi_k}$ in the equation 
(1). On the other hand, as suggested 
by observation [4] we need $\mid n_s-1\mid <0.1$, so $\mid kt_0\mid<0.1$. Introducing 
dimensionless $T$ and $R$ as $t_0=10^T/M_p$ and $\rho=M_p^4/10^R$ one get the inequality
$$
R-2T>2+\lg(8\pi/3)\sim 2.92.
$$
This inequality represent restriction to possible values of $t_0$ and initial energy density 
$\rho$. One can see that if  $t_0\sim 10^{-35}$s then $\rho<10^{-19}M_p^4$ so we have not normal 
initial conditions for the chaotic inflation.

\section{Density perturbations.}

As we have seen the Darboux transformation allows one to obtain a set of potentials 
which lead to the flat (or almost flat) spectrum. But we should distinguish between the 
spectrum of the scalar field and 
the spectrum of density perturbations~\footnote{I'd like to thank prof. A. Linde who drew 
my attention  to this circumstance.}. An investigation of this question by Lyth [7], 
Brandenberger [8] 
and others suggests that even though the scalar field perturbations with flat spectrum can 
be generated by this effect, it does not lead to perturbations of metric with flat spectrum. 
In particularly, Brandenberger in [8] has considered the simplest potential which lead to the scale 
factor $a(t)\sim t^p$. For extremely slow contraction with $p\sim 0$ he get a blue spectrum 
with index $n\sim 3$. It is interesting to consider more general class of potentials and 
we'll do it in this section. 

In the case of general position one can't solve the Einstein equations exactly to 
find the scale factor $a(t)$ but using the Darboux transformations we can construct 
a rich set of exact soluble potentials so this method it is interesting itself. 

Let consider the Einstein equations (in the units with $c=8\pi G=1$)
$$
\ddot\phi+3H\dot\phi+V'=0,\qquad 
H^2=\frac{1}{3}\left(\frac{1}{2}{\dot\phi}^2+V(\phi)\right),
\eqno(20)
$$
so ${\dot\phi}^2=-2\dot H$. Introducing $\psi(t)=a(t)^3$ one get $\ddot\psi=3V\psi$. 
If we add positive cosmological term $\Lambda>0$ then this equation can be written as 
the Schr\"odinger equation
$$
\ddot\psi=\left(u+\lambda^2\right)\psi,
\eqno(21)
$$
where $u(t)=3V(\phi(t))$, $\lambda^2=3\Lambda$. We denote the scale factor and the 
Hubble "constant" in the universe with $\lambda=0$ as $a=a(t)=\psi^{1/3}$, $H=H(t)$ 
whereas the same in the universe with $\lambda\ne 0$ will be $\tilde a={\tilde\psi}^{1/3}$ 
and $\tilde H$.

We suppose  that in our universe the $\lambda^2=0$. Some researchers are suggested that 
the reason of the recently discovered accelerated expansion of the universe is the 
nonzero cosmological term. 
Indeed, it is easy to formulate usual  
inflationary theory to obtain this accelerated expansion. 
 To do it one can add a small constant term $V_0>0$ to the potential 
$V(\phi)$ [9]~\footnote{Cosmology with negative potentials were considered in [10].}.
If cosmological constant $V_0\sim 10^{-120}$ (in Planck units) then one 
get the present accelaration [10], but is not terrifically because it is not clear 
why should $V_0$ be so small? This approach return us to the old mystery of 
vacuum energy [11] and this is why we suppose that $\lambda^2=0$ in our universe. 
The solutions of (20) with  nonzero cosmological  term will be necessary for the 
Darboux transformation (see below).

The Darboux transformation for the (21) has the form 
$$
\psi\to \psi^{(1)}=\left(\sigma-{\tilde{\sigma}}\right)\psi,\qquad 
u\to u^{(1)}=u-2\dot{\tilde{\sigma}},
$$
where $\sigma=d\log\psi/dt$, ${\tilde{\sigma}}=d\log{\tilde{\psi}}/dt$. One get
$$
H^{(1)}=-{\tilde H}+\frac{\lambda^2}{9({\tilde H}-H)},
\eqno(22)
$$
therefore the dressed scale factor $a^{(1)}(t)$ has the form
$$
a^{(1)}=\frac{const}{\tilde a}\exp\left[-\frac{\lambda^2}{9}\int dt 
\left(\frac{d}{dt}\log\frac{\tilde a}{a}\right)^{-1}\right].
\eqno(23)
$$
At last, dressed  potential $V^{(1)}(\phi^{(1)})$ can be written in 
parametrical form,
$$
\begin{array}{l}
\displaystyle{
\phi^{(1)}(t)=\phi^{(1)}_0-\sqrt{\frac{2}{3}}\int dt\sqrt{\frac{\lambda^4}{9(H-\tilde H)^2}+
\lambda^2\frac{H+\tilde H}{H-\tilde H}+3\dot{\tilde H}},}\\
V^{(1)}(t)=V(t)-2\dot{\tilde H},
\end{array}
\eqno(24)
$$
where $\phi^{(1)}_0$ is constant. 

This method of  construction of exact soluble potentials is valid both for usual 
inflation cosmology and for the effective field theory of the Ekpyrotic/Cyclic Universe.  
In the last case the scalar field $\phi$ represents the position of the bulk brane in the fifth 
dimensions.  
Following Brandenberger one consider the scale factor $a(t)\sim (-t)^p$ (time is negative). 
If $\lambda^2=0$ then 
$$
u(t)=\frac{3p(3p-1)}{t^2}.
$$
The solution of (21) with $\lambda^2\ne 0$ has the form 
$$
\tilde\psi=\sqrt{-t}\left(c_1K_{\nu}(-\lambda t)+c_2 I_{\nu}(-\lambda t)\right),
\eqno(25)
$$
where $\nu=\pm(1-6p)/2$, $c_1>0$ and $c_2>0$ are constants, $K_{\nu}$ and $I_{\nu}$ are 
modified  Bessel function (if $\Lambda<0$ then the solution of the (21) will be 
expressed via Hankel  functions but the final result will be the same), 
$t<0$ and we choose $\nu>-1$ to obtain real and positive solution (25). For the 
small $\lambda t$ one get 
$$
\tilde\psi=\frac{c_1}{\sqrt{2\lambda}}\Gamma(\nu)\left(\frac{-\lambda t}{2}\right)^{-\nu-1/2}+
c_2\sqrt{\frac{2}{\lambda}}\Gamma(\nu+1)\left(\frac{-\lambda t}{2}\right)^{-\nu+1/2}.
\eqno(26)
$$
Using (23), (26) and well known relations between modified Bessel functions and it's derivatives 
[12] we have  $a^{(1)}\sim (-t)^{p^{(1)}}$. The power $p^{(1)}$ is depend on initial power 
$p$. For example, if $1/3<p<1/2$, then $p^{(1)}=p-1/3$, so $0<p^{(1)}<1/6$, and we have 
blue spectrum with index 
$$
n=1+\frac{2}{1-p^{(1)}}=1+\frac{6}{4-3p}.
$$
If $p\to 1/3$ then we get the extremely slow contraction and the spectrum with index 
$n\sim 3$. 

Another values of $p^{(1)}$ can be obtained starting out from the another 
choices of $p$.   The general conclusion is that all new potentials leads to the 
old result of Brandenberger. In other words we have not the flat spectrum of 
perturbations of metric.

\section{Conclusion.}

Up to know, Darboux transformations didn't used in cosmology (by contrast with quantum mechanics 
and the theory of integrable PDE). In this paper we are used only simple example of  
 Darboux transformation. 
One can construct infinite set of potentials which lead to the flat spectrum with arbitrary 
multiple precision. We can do it without any brane-string physics  and 
we don't need consider field theory with power-low potential  when $n>40$. 

On the other hand, it is valid only for the spectrum of the scalar field and, to all appearances, 
don't valid for the spectrum of density perturbations. Using Darboux transformations, one can 
construct infinite set exact soluble cosmologies with flat spectrum of the scalar field 
perturbations and, at the same time, with "essentially blue" spectrum of perturbations 
of metric. 
$$
{}
$$
{\em Acknowledgements.} The author are grateful to Andrei Linde for useful comments. 
This work was supported by the Grant of Education Department of the Russian Federation, 
 No. E00-3.1-383.
$$
{}
$$
\centerline{\bf References}
\noindent
\begin{enumerate}
\item  R. Kallosh, L. Kofman, and A. Linde, \rm\, Phys. Rev. D {\bf 64}, 123523 (2001).
\item J. Khouru, B.A. Ovrut, P.J. Steinhardt and N. Turok,\rm\, Phys. Rev. D {\bf 64}, 123522
(2001).
\item P.J. Steinhardt and N. Turok,\rm\,"A Cyclic Model of the Universe", [hep-th/0111030] (2001).  
\item  A.H. Jaffe et al., \rm\, Phys. Rev. Lett. {\bf 86}, 3475 (2001).
\item  V.B. Matveev and M.A. Salle.\rm\,  Darboux Transformation and
    Solitons. Berlin--Heidelberg: Springer Verlag, 1991.
\item M. Bordag and A. Yurov,\rm\, "Spontaneous Symmetry Breaking and Reflectionless Scattering 
Data", [hep-th/0206199] (2002).
\item D. H. Lyth and D. Wands,\rm\, Phys. Lett. B {\bf 524} 524, 5-14 (2002) 
[hep-ph/0110002].
\item R. Branderberger and F. Finelli, \rm\, "On the Spectrum of Fluctuations in the Field 
Theory of the Ekpyrotic Universe", JHEP {\bf 0111}, 056 (2001) [hep-th/0109004].
\item A. Linde, \rm\, "Inflationary Theory Versus Ekpyrotic/Cyclic Scenario",  
[hep-th/0205259] SU-ITP-02-25 (2002).
\item G. Felder, A. Frolov, L. Kofman and A. Linde, \rm\, "Cosmology With Negative Potentials", 
CITA-2002-04 SU-ITP-02/05 [hep-th/0202017] (2002).
\item A.D. Dolgov, "Mystery of Vacuum Energy or Rise and Fall of Cosmological Constant", \rm\,
[hep-ph/0203245] v1 (2002).
\item "Handbook of Mathematical Functions with Formulas, Graphs and Mathematical Tables",  edited 
by M. Abramowitz and I.A. Stegun, National Bureau of standards applied mathematics series 55, 1964.

\end{enumerate}

\end{document}